# Liquid crystal elastomer coatings with programmed response of surface profile

Greta Babakhanova[1,2], Taras Turiv[1,2], Yubing Guo[1,2], Matthew Hendrikx[3], Qi-Huo Wei[1,2], Albert P.H.J. Schenning[3,4], Dirk J. Broer[3,4] & Oleg D. Lavrentovich[1,2,5*]

[1]*Liquid Crystal Institute, Kent State University, Kent, OH 44242, USA*

[2]*Chemical Physics Interdisciplinary Program, Kent State University, Kent, OH 44242, USA*

[3]*Functional Organic Materials and Devices, Department of Chemical Engineering and Chemistry, Eindhoven University of Technology, 5612 AZAE, Eindhoven, The Netherlands*

[4]*Institute for Complex Molecular Systems, Eindhoven University of Technology, P.O. Box 513, 5600 MB, Eindhoven, The Netherlands*

[5]*Department of Physics, Kent State University, Kent, OH 44242, USA*

**Stimuli-responsive liquid crystal elastomers (LCEs) with a strong coupling of orientational molecular order and rubber-like elasticity, show a great potential as working elements in soft robotics, sensing, transport and propulsion systems. We demonstrate a dynamic thermal control of the surface topography of LCE coatings achieved through pre-designed patterns of in-plane molecular orientation. These patterns determine whether the LCE coating develops elevations, depressions, or in-plane deformations. The deterministic dependence of the out-of-plane dynamic surface profile on the in-plane orientational pattern is explained by activation forces. These forces are caused by two factors: (i) stretching-contraction of the polymer networks driven by temperature; (ii) spatially varying orientation of the LCE. The activation force concept brings the responsive LCEs into the domain of active matter. The demonstrated relationship can be used to design programmable coatings with functionalities that mimic biological tissues such as skin.**

Biological systems demonstrate the importance of an intelligent design of interfacial tissues, or "skins", in a variety of aspects, including self-cleaning and self-healing, suction, friction, touch perception, stretching and contraction, transport of matter along and across the coating. Very recent research breakthroughs[1,2] demonstrated two key factors controlling cell dynamics in biological tissues such as epithelium: (i) orientational order of cells, and (ii) presence of topological defects in this orientational order. Activity of individual cells combined with the spatially-varying order leads to compressive-dilative stresses in the tissue that facilitate in-plane and out-of-plane displacements. In this work, we demonstrate that similar displacements and changes in the surface profile can be achieved in the temperature-activated artificial nematic liquid crystal elastomer (LCE) coatings with patterns of orientational order pre-imposed in the plane of the coating during polymerization. The out-of-plane and in-plane displacements are deterministically related to the type of deformations (splay or bend) of the orientational order pre-inscribed into the LCE.

An LCE is an anisotropic rubber, as it is formed by cross-linked polymeric chains with rigid rod-like mesogenic segments in the main chain and attached as side branches; these mesogenic units are similar to the molecules forming low-molecular weight liquid crystals [3,4]. The mesogenic moieties in the nematic state of an LCE are oriented along a certain nonpolar direction called the director $\hat{\mathbf{n}} \equiv -\hat{\mathbf{n}}$. Cross-linked polymeric chains are structurally anisotropic because of their coupling to the orientational order. The coupling enables mechanical response of LCEs to external factors such as temperature. For example, upon heating, a uniformly aligned LCE strip contracts along the director and expands in the perpendicular directions, since the orientational order weakens and the cross-linked polymer network becomes more isotropic [3,5]. Such a uniform LCE strip behaves as an artificial muscle [6]. Recent research unraveled even more exciting effects when the director changes in space, $\hat{\mathbf{n}}(\mathbf{r}) \neq const$. Thin LCE films with in-plane director patterns develop 3D shape changes with non-trivial mean and Gaussian curvatures when exposed to thermal activation [4,7-11], while director deformations across the film trigger wave-like shape changes and locomotion when activated by light illumination[12]. In a parallel vein, there is a tremendous progress in exploiting LCE coatings in which one surface is attached to the substrate and the other is free [13,14]. When illuminated with light, photoresponsive coatings with a misaligned director develop random spike-like topographies [15], while periodic elevations and groves can be produced by using cholesteric "fingerprint" textures [16], or periodic stripe arrays [17]. The challenge is in finding an approach by which the change of the topography of the coating or its stretching/contraction can be deterministically pre-programmed. In this work, we demonstrate such an approach, by establishing a deterministic relationship between the splay and bend deformations of the director patterns inscribed in the plane of an initially flat LCE coating at the stage of preparation, on one hand, and spatially-varying surface topography, in-plane and out-of-plane shears and deformations of this coating under thermal addressing, on the other hand.

The director patterns with pre-designed splay and bend of the director are imposed onto the LCE by the plasmonic photoalignment technique [18] (see Supplement Information). Two parallel glass plates separated by a distance of 5 μm coated with photosensitive azodye molecules are irradiated by a light beam that passes through a plasmonic photomask with a patterned array of elongated nanoapertures. The photomask imparts spatially-varied linear polarization onto the transmitted light beam[18]. Under irradiation, the azodye molecules reorient their long axes perpendicularly to the local light polarization [18]. When the liquid crystal in its monomeric state fills the gap between the two glass plates, the patterned surface orientation of the azodye molecules establishes the director pattern in the bulk. The mixture is then photopolymerized to obtain the LCE with the desired pattern of the director. The director is modulated in the $(x, y)$ plane of the film and remains parallel to the bounding plates; there is no change of the director through the thickness of the cell, as the top and bottom plate are photo-patterned in the same way. One of the plates is removed to obtain the LCE coating. At the stage of preparation, the coating is practically flat.

Figures 1a, 2a, 3a show the director patterns imprinted into the LCE coatings with arrays of topological defects of integer and semi-integer strength. The director field is of the form

$$\hat{\mathbf{n}} = (n_x, n_y, 0) = (\cos\alpha, \sin\alpha, 0) \tag{1}$$

where $\alpha(x, y) = \sum_i m_i \arctan\left(\dfrac{y - y_{0i}}{x - x_{0i}}\right) + \varphi_0$, $m_i = \pm 1/2, \pm 1$ indicates the strength of the defect, $x_{0i}$ and $y_{0i}$ are the coordinates of the core of the defect and $\varphi_0$ is the constant phase specifying the prevailing type of deformations and orientation of the defect structures. The +1 defects are of a radial type, carrying deformations of pure splay, when $\varphi_0 = 0$ (Fig.1a), and of a circular geometry, carrying pure bend, when $\varphi_0 = \pi/2$ (Fig.2a). The -1 defects in both cases exhibits four alternating regions of splay and bend.

The LCE coating with the pre-inscribed director pattern is practically flat at the room temperature. The surface profile is established by Digital Holographic Microscopy (DHM). When the coating is heated above the glass transition temperature (~ 50°C), its surface topography shows a remarkably robust and reproducible change determined by the specific form of director distortions. Namely, pure splay regions associated with $m = +1$ and $\varphi_0 = 0$ defects (Fig.1a), produce depressions (Fig.1b,c,d); the material moves away from the center of the defect along the radial directions. In contrast, pure bend regions associated with $m = +1$ and $\varphi_0 = \pi/2$ defects in Fig.2a, produce pronounced hills (Fig.2b,c,d). The amplitude of thickness variation is about 300-400 nm, representing a substantial fraction of the total thickness ($5\,\mu m$). The heating-induced non-flat profile around $m = -1$ defects is more complicated, exhibiting four ridges and four valleys (Fig. 1b,c and Fig. 2b,c). The dynamics of surface topography is completely reversible when the temperature is cycled between the room temperature and the maximum temperature of around 120˚C. The reversibility is lost if the material is heated into an isotropic state with no orientation order (at about 150˚C).

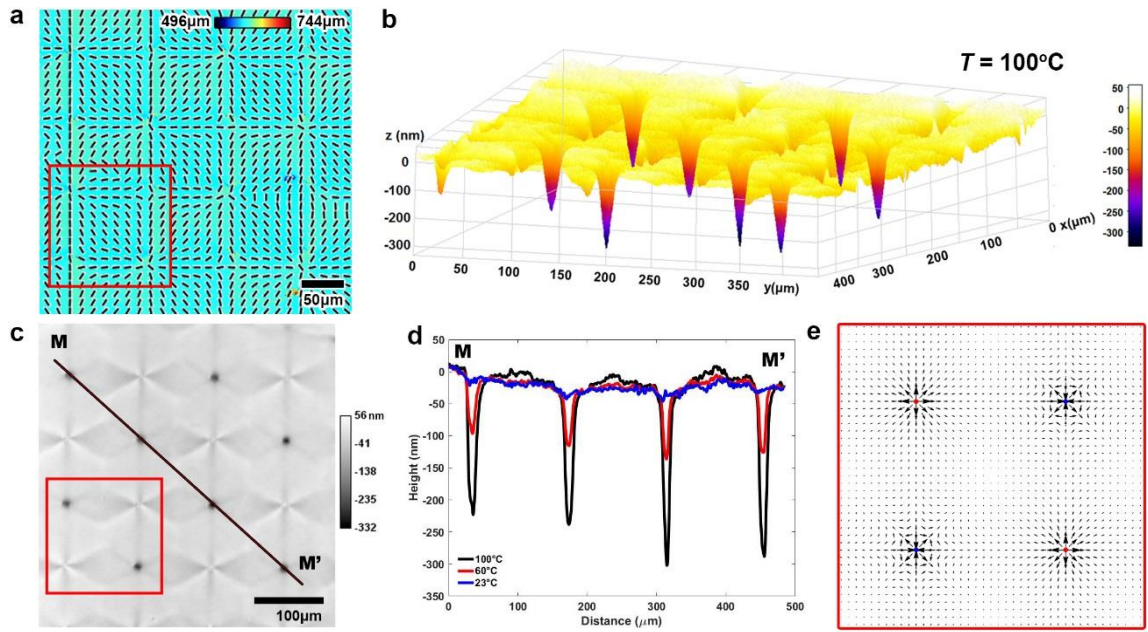

**Figure 1 Surface topography deformations produced by +1 defect with splay deformation. a,** PolScope image of LCE which maps the optical retardance and the director orientation. **b,** 3D image of surface topography of LCE at 100˚C observed using DHM. **c,** DHM image of the LCE surface at 100˚C used to extract the surface topography. **d,** Surface profiles along line *MM'* in (c) at 23˚C, 60˚C, 100˚C. **e,** Activation force density **f** map calculated for region outlined by a red box in (a,c).

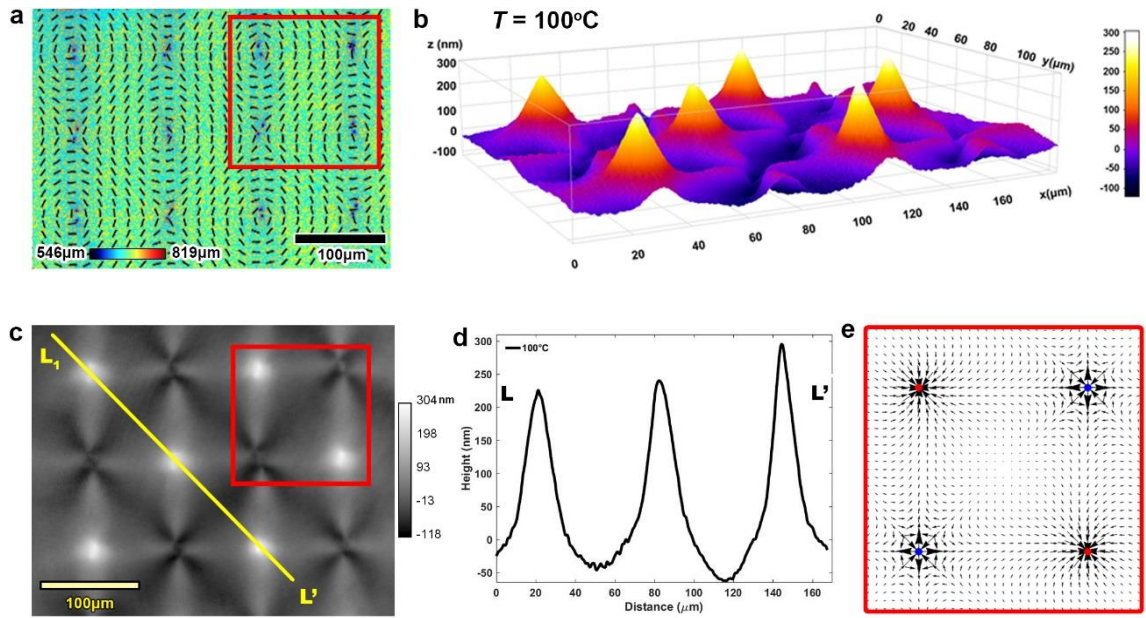

**Figure 2 Surface topography deformations produced by +1 defect with bend deformation. a,** PolScope image of LCE which maps the optical retardance and the director orientation. **b,** 3D image of surface topography of LCE at 100˚C observed using DHM. **c,** DHM image of the LCE surface at 100˚C used to extract the surface topography. **d,** Surface profiles along line *LL'* in (c) at 100˚C. **e,** Activation force density **f** map calculated for region outlined by a red box in (a,c).

Similar relationships between the director patterns and surface topography are observed in patterns with $m=+1/2$ and $m=-1/2$ defects (Fig 3). The defects $m=+1/2$ in Fig.3 are of a polar symmetry with one bend regions and one splay region; the $m=-1/2$ defects exhibit three regions of bend and three regions of splay each (Fig. 3a). The corresponding surface profile is comprised of a single elevation/depression pair in $m=+1/2$ case (Fig. 3b) and three elevations/depressions around the $m=-1/2$ defects, as expected from the symmetry of the defects. There is a new unique feature of the $m=+1/2$ defects, not observed neither for $m=-1/2$ nor for $m=\pm 1$ defects. Namely, upon heating, both the core and the elevation/depression pair associated with the $m=+1/2$ defect shift along the vector directed from the bend region towards the splay region (Fig.3c,e,f). Figures 3e and 3f show the temperature effect on the distance between $m=\pm 1/2$ defect pairs A and B, where the origin (0,0) is their initial separation distance at 30˚C. The shift of $m=+1/2$ defect core is fully reversible in the heating-cooling cycles, provided the maximum temperature does not exceed 120˚C.

The displacement of the elevation/depression pair is observed using DHM (Fig.3c), by tracing the temperature dependent topography along *TT'* line which crosses three defect cores: -1/2, +1/2 and -1/2 (Fig.3b). The inset in Fig.3c demonstrates that on heating the LCE by 40˚C, the shift $\Delta y$ of elevation is about $5\ \mu m$. The displacement of the depression is about $3\ \mu m$. The associated shift of the $m=+1/2$ defect core is about $3\ \mu m$ as observed under PolScope (Fig.3e,f). The core is the central region of the defect at which the orientational order is strongly diminished. Its position is thus readily visualized by PolScope that maps the local optical retardance since the optical retardance at the core is much lower than in the rest of the sample (Fig.3a,e) [19].

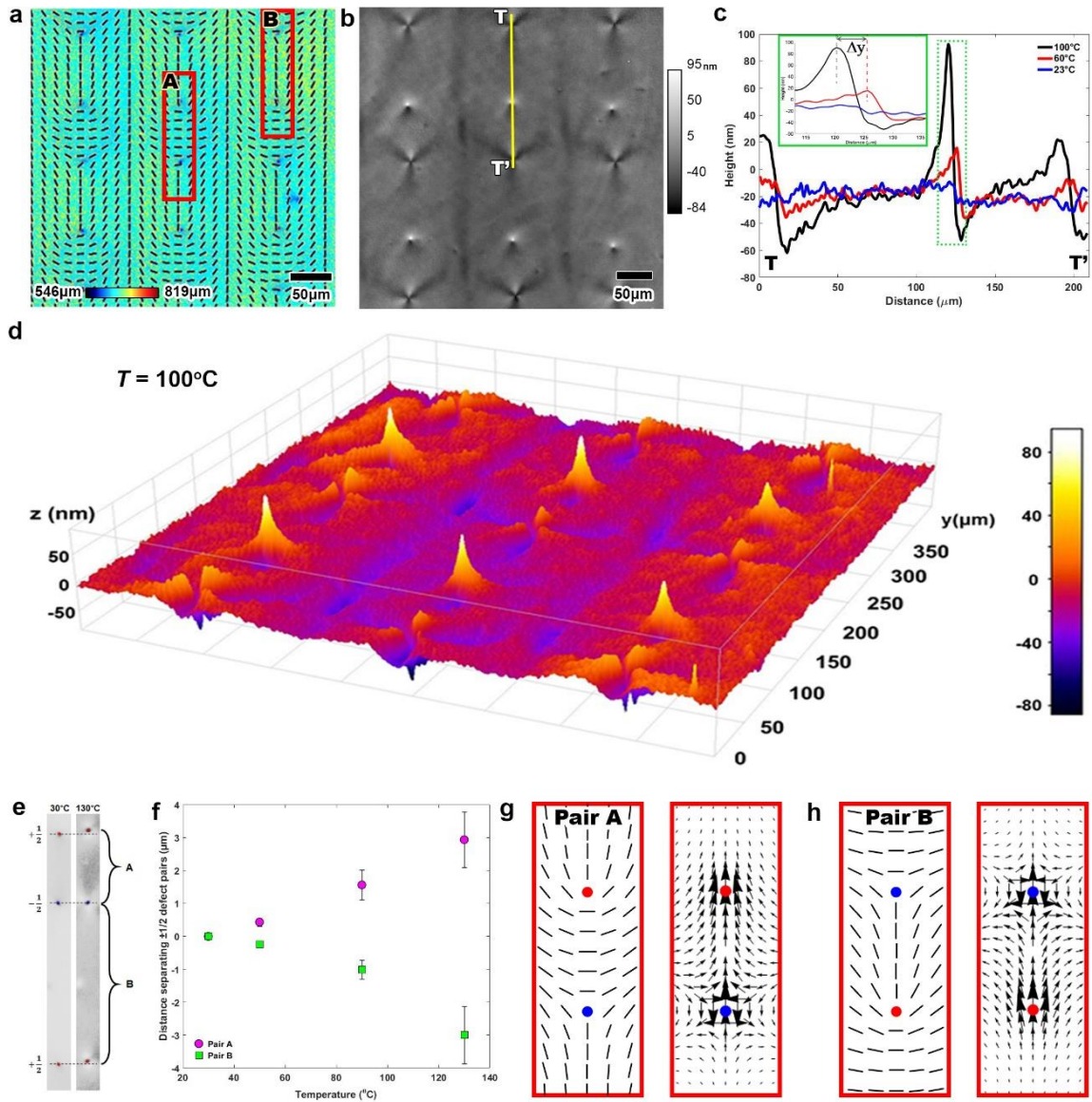

**Figure 3 Surface topography deformations produced by half-integer defects. a,** PolScope image of LCE which maps the optical retardance and the director orientation, red boxes encompass the half-integer defects separated by director field configurations A and B. **b,** DHM image of the LCE surface at 100°C used to extract the surface profile along line *TT'*. **c,** Surface profiles along line *TT'* in (b) at 23°C, 60°C and 100°C. **d,** 3D image of surface topography of LCE at 100°C observed using DHM. **e,** Grayscale PolScope images taken at 30°C and 130°C showing the displacement of the half-integer defects as a function of temperature and director configuration that separates the defects. **f,** Plot showing the distance separating ±1/2 defects as a function of temperature and director configurations A and B. **g,h,** Activation force density **f** maps calculated for director configurations for Pair A and Pair B respectively.

The correlation splay-depression and bend-elevation is observed not only in arrays with integer-strength topological defects but also in defect-free patterns of alternating splay and bend. As an example, Fig.4 shows the response of an LCE coating with the director field $\hat{\mathbf{n}} = (n_x, n_y, n_z) = (|\cos\beta|, \sin\beta, 0)$, where $\beta(y) = \pi y / P$ and $P$ is half the period. The profile along the line *NN'* in the DHM texture (Fig 4a) changes from being flat at room temperature to strongly modulated at elevated temperatures, with alternating sharp valleys at the locations of a maximum splay and ridges at locations with a prevailing bend.

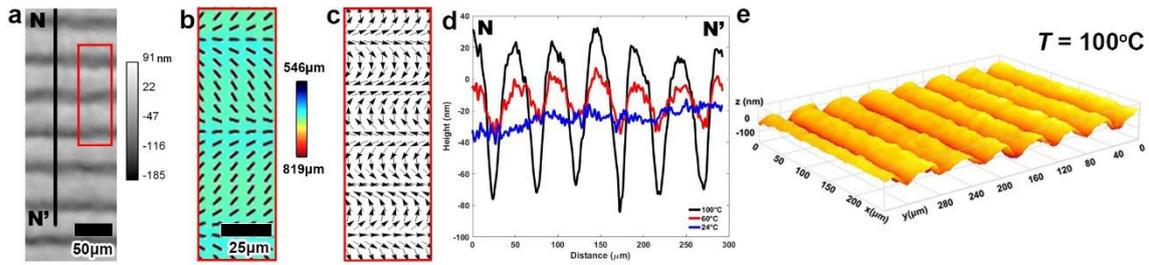

**Figure 4 Surface topography deformations produced by splay director deformations. a,** DHM image at 100˚C used to extract the surface profile along line *NN'*. **b,** PolScope image of LCE which maps the optical retardance and the director orientation. **c,** Force density map calculated for region in (a,b). **d,** Surface profiles along line *NN'* in (a) at 24˚C, 60˚C and 100˚C. **e,** 3D image of surface topography of LCE at 100˚C observed using DHM.

The experiments above demonstrate clearly that the heating-induced dynamic profile of the LCE coating is defined deterministically by the type of director deformations pre-programmed in the plane of the initially flat sample.  For example, in Fig.1, 2, and 4, pure splay causes depressions while pure bend causes elevations. The relationship is more complicated around $m = +1/2$ defects in Fig.3, since the vector connecting a depression to an elevation is directed from bend to splay region (Fig. 3b). The underlying mechanism can be understood by considering the microscopic response of the LCE to a changing temperature.

Orientational order is coupled to mechanical deformations of LCEs, because of the cross-linking of the polymer network. This coupling results in an anisotropic structure of the network characterized by the so-called step length tensor [3] $l_{ij} = l_\perp \delta_{ij} + (l_\parallel - l_\perp) n_i n_j$. The step length $l$ characterizing the polymer segments connecting cross-linking points is different when measured along $\hat{\mathbf{n}}$ ($l_\parallel$) and perpendicularly ($l_\perp$) to $\hat{\mathbf{n}}$. For $l_\parallel > l_\perp$, the spatial distribution of the step lengths can be represented by a prolate ellipsoid elongated along $\hat{\mathbf{n}}$ (Fig.5a). If the temperature is raised and the orientational order weakens, the distribution becomes more spherical, i.e., the ellipsoid shrinks along $\hat{\mathbf{n}}$ and expands in two perpendicular directions. Once the nematic order is melted, the ellipsoid becomes a sphere, $l_\parallel^{iso} = l_\perp^{iso} = \bar{l}$ (Fig.5b).

Morphing of the step length ellipsoid caused by the temperature can be modelled by a force dipole, with two point forces of equal amplitude $F$ directed from the poles of the ellipsoid towards its center (Fig.5a). For an LCE of a constant volume and small anisotropy, $l_\perp - l_\parallel \ll l_\perp, l_\parallel$, this amplitude can be estimated as $F \sim \mu \bar{l} (l_\perp - l_\parallel)$, where $\mu$ is the shear modulus of the LCE, on the order of $(10^4 - 10^5)$ J/m$^3$ [20]. Whenever the director field of the LCE changes in space, so do the local axes of the ellipsoids (Fig. 5c,d). The spatial gradients of the step-length tensor produce a vector quantity with the components $f_i = \alpha \partial_j n_i n_j$, which can also be written in the equivalent invariant form as

$$\mathbf{f} = \alpha (\hat{\mathbf{n}} \operatorname{div} \hat{\mathbf{n}} - \hat{\mathbf{n}} \times \operatorname{curl} \hat{\mathbf{n}}), \tag{2}$$

where $\alpha \sim F / \bar{l}^2 \sim \mu (l_\perp - l_\parallel) / \bar{l}$ is introduced as an activation parameter that describes the local elastic response to the changing temperature; for the sake of simplicity, the estimated value of $\alpha$ corresponds to the complete melting of orientational order. Note that $\alpha$ depends on the dimensionless anisotropy $(l_\perp - l_\parallel) / \bar{l}$ rather than on the absolute values of the step lengths, which stresses a universal character of the elastic response of LCEs with little dependence on the concrete microscopic details [21]. In the order of magnitude, with $(l_\perp - l_\parallel) / \bar{l} \sim 0.1$, one expects $|\alpha| \sim (10^3 - 10^4)$ J/m$^3$. When the temperature of an LCE with $l_\parallel > l_\perp$ increases and the long axes of the polymer ellipsoids shrink, then $\alpha > 0$; in the case of cooling, $\alpha < 0$.

With $\alpha$ defined as above, the vector $\mathbf{f}$ represents a spatially varying activation force density that controls the elastic response of an LCE with a non-uniform director $\hat{\mathbf{n}}(\mathbf{r}) \neq const$ to the external factors such as heating that bring the system out of equilibrium. The occurrence of the force $\mathbf{f}$ is illustrated in Fig.5c,d for the case of pure splay and pure bend, respectively. For example, in the case of bend, Fig.5d, the point forces of the two neighbouring shrinking ellipsoids that are tilted with respect to each other, produce a net force density $\mathbf{f}$ along the radius of curvature of the director $\hat{\mathbf{n}}(\mathbf{r})$.

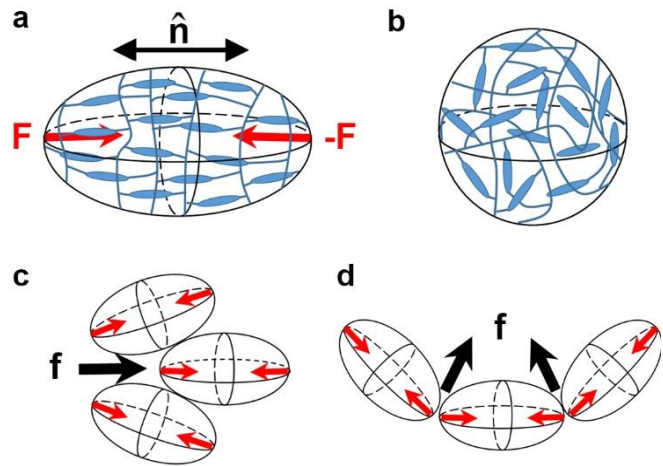

**Figure 5 Ellipsoid of polymer network conformation and the occurrence of the activation force. a,** Prolate ellipsoid of polymer network conformations in the nematic phase; the long axis is along the director $\hat{\mathbf{n}}$ ; during heating, the ellipsoid shrinks along the long axis; in the isotropic phase, it becomes a sphere, as shown in part **(b);** the shrinking ellipsoid is modelled by a pair of forces $\mathbf{F}$. **c,** Activation force density $\mathbf{f}$ produced by contracting ellipsoids in the geometry of pure splay. **d,** Activation force density $\mathbf{f}$ produced by contracting ellipsoids in the geometry of pure bend.

The activation coefficient $\alpha$ and the vector $f_i = \alpha \partial_j n_i n_j$ are similar to the activity coefficient and active current introduced by Simha and Ramaswamy [22] in the description of flowing active matter with elementary swimming units representing force dipoles of either puller of pusher type; the former are similar to the shrinking LCE ellipsoids (Fig.5a). Furthermore, the factor $\hat{\mathbf{n}}\,\mathrm{div}\,\hat{\mathbf{n}} - \hat{\mathbf{n}} \times \mathrm{curl}\,\hat{\mathbf{n}}$ is identical to the gradient expression of the flexoelectric polarization in a spatially-nonuniform liquid crystals in which the molecular structure is of quadrupolar symmetry [23]. It is also the same as the gradient part of the active force considered by Green, Toner and Vitelli [24] for distorted incompressible active nematic fluids. All these similarities bring under one umbrella rather different phenomena, flow of active fluids, flexoelectric polarization of a distorted nematic liquid crystal and stimuli-responsive LCEs. The connection is not surprising as in all cases the symmetry of elementary units (ellipsoids of shrinking-expanding polymer networks in the case of the LCE) is quadrupolar and the alignment direction of these units varies in space. Note that the force $\mathbf{f}$ in equation (2) is not uniquely associated with the heating/cooling and can be used to describe the effect of other agents, such as light irradiation, humidity change, etc.

The spatial distribution of the activation force $\mathbf{f}$ acting in pre-programmed director patterns (Fig.1e,2e,3g,h) qualitatively explains the observed deterministic relationships such as splay-depression and bend-elevation. Consider first the case of $m=+1$ defects of a radial type with splay (Fig.1, 5c) and a circular type with bend (Fig.2, 5d). The force $\mathbf{f}$ is directed away from the center of the radial defect in Fig.1e and towards the center of the circular defect in Fig. 2e. This force transports the matter, thus the LCE coating upon heating becomes depleted around the radial $m=+1$ defect with pure splay and elevated in the region of the circular $m=+1$ defect of pure bend. In the case of $m=-1$ defects, the force produces four ridges and four valleys, by converging in the bend regions and diverging in the splay regions. For example, the spatial map of the activation force shown in Fig.1e predicts valleys along the horizontal and vertical directions, and the ridges that are at 45 degrees to the valleys, which is what is seen in the experiment (Fig.1b,c,d).

The activation force $\mathbf{f}$ field also helps to understand why the cores of $m=+1/2$ upon heating move towards the splay region: the angular distribution of the force around the $m=+1/2$ defect core is not symmetric, breaking the fore-aft symmetry with a nonzero net force directed towards the splay (tail) region. This effect is yet another demonstration of a deep analogy between the nonequilibrium behavior of responsive LCE and active matter such as arrays of vibrating rods [25], living cell cultures [2,26], bacterial colonies [27] and microtubules powered by kinesin motors [28]. In all these systems, $m=+1/2$, unlike their $m=-1/2,\pm1$ counterpart, show a net propulsion in out-of-equilibrium dynamics, either in the direction of bend or the splay, depending on whether the active units are extensile or contractile; these types differ in the sign of the activity coefficient $\alpha$. The heated LCE corresponds to a contractile case, $\alpha > 0$; an extensible version with $\alpha < 0$ could be manufactured by polymerising the nematic LCE at elevated temperatures and then cooling it down. The map of the activating force for $m = +1/2$ suggests that the material is pulled along the vector that is directed from the bend region towards the splay region. An exact analytical description of the resulting profile of the LCE coating is not easy to construct, since the solution should account for mass conservation, dynamic coupling of the director field and rubber elasticity to the material transport, different boundary conditions at the two interfaces, etc.

To conclude, we demonstrated that the dynamic surface profile of the LCE coatings activated by temperature can be pre-programmed deterministically by inscribing in-plane patterns of orientational order into the initially flat elastomer at the stage of preparation. Deformations such as splay, bend and their combinations cause different response of the LCE coating, triggering topography changes with local elevations, depressions and in-plane shifts. The mechanism of the effect is explained by introducing an activation force that makes the description of dynamically addressed LCEs similar to that of active matter. The proposed approach implies that the activated LCEs can serve as a well-controlled experimental model of active matter, which is a very welcome addition to this rapidly developing field. From the practical point of view, the ability to control 3D shape changes through 2D inscribed patterns of orientational order could be used to produce a variety of useful materials and devices, such as controllable origami, coatings and films with dynamic hydrophobicity/hydrophilicity patterns, coatings that could move microparticles in space according to the underlying topography of their surface, etc.

Supplementary Information is available in the online version of the paper.

Reprints and permissions information is available at www.nature.com/reprints

**Acknowledgements** The work was supported by the NSF grant DMR-1507637, the Netherlands Organization for Scientific Research (NWO; TOP PUNT grant 10018944) and the European Research Council (Vibrate ERC, grant 669991). The SEM data were obtained at the Characterization facility at the Liquid Crystal Institute, Kent State University. The authors thank R. Green, D. Liu, S. Paladugu, S. V. Shiyanovskii, J. Toner, and V. Vitelli for fruitful discussions. ODL and GB are thankful for the kind hospitality of Eindhoven University of Technology for the hospitality during their visit to Drs. Broer and Schenning laboratories where preparation and DHM characterization of the LCE coatings were performed.

**Author Contributions** G.B. and O.D.L. designed the experiments and wrote the manuscript with the input from all co-authors; G.B. analysed the data; O.D.L. developed the theoretical model; T.T. assisted with the force field calculations and substrate preparation; Y.G. and Q.W. produced the plasmonic metamasks; M.H. assisted with DHM experiments. A.P.H.J.S. and D.J.B. provided the LCE materials, advised on the materials properties and techniques of LCE preparation and participated in discussing and interpreting the results; O.D.L. supervised the project. All authors contributed to the editing of the manuscript.

**Author Information** The authors declare no competing financial interests. Correspondence and requests for materials should be addressed to O.D. Lavrentovich (olavrent@kent.edu).